\documentclass[12 pt]{article}   

\newcommand{\equ}[1]{Eq.\,(\ref{#1})}
\newcommand{\eqs}[1]{Eqs.\,(\ref{#1})}

\newcommand{\non}{\nonumber}
\newcommand{\gsim}{\;\rlap{\lower 3.5 pt \hbox{$\mathchar \sim$}} \raise 1pt
 \hbox {$>$}\;}
\newcommand{\lsim}{\;\rlap{\lower 3.5 pt \hbox{$\mathchar \sim$}} \raise 1pt
 \hbox {$<$}\;}
\newcommand{\smallw}{{\scriptscriptstyle W}} %

\newcommand{\mw}{M_\smallw}

\def\zp#1#2#3{{\it Z. Phys. }{\bf C#1~}(19#2)~#3}
\def\prl#1#2#3{{\it Phys. Rev. Lett. }{\bf #1~}(19#2)~#3}

\def\pr#1#2#3{{\em Phys. Rev. }{\bf D#1~}(19#2)~#3}
\def\np#1#2#3{{\em Nucl. Phys. }{\bf B#1~}(19#2)~#3}
\def\ap#1#2#3{{\it Ann. Phys.} (NY) #1 (19#2) #3}
\def\cmp#1#2#3{{\it Comm. Math. Phys.} #1 (19#2) #3}

\newcommand{\be}{\begin{equation}}
\newcommand{\ee}{\end{equation}}
\newcommand{\een}{\end{subequations}}
\newcommand{\ben}{\begin{subequations}}
\newcommand{\beq}{\begin{eqalignno}}
\newcommand{\eeq}{\end{eqalignno}}
\newcommand{\bea}{\begin{eqnarray}}
\newcommand{\eea}{\end{eqnarray}}

\def\h{\hat}

\def \ms   {\overline{\mbox{MS}}}

\begin{document}              
\begin{titlepage}
\begin{flushright}
        \small
        TUM-HEP-332/98\\
        MPI-PhT/98-86\\
        hep-ph/9811470\\
        October 1998
\end{flushright}

\begin{center}
\vspace{1cm}
{\Large\bf Fermion  mixing  renormalization    and \\ gauge invariance}

\vspace{0.5cm}
\renewcommand{\thefootnote}{\fnsymbol{footnote}}
{\bf           P.~Gambino$^a$,
                        P.A.~Grassi$^b$, and 
F. Madricardo$^b$}
\setcounter{footnote}{0}
\vspace{.8cm}

{\it
        $^a$ Technische Universit{\"a}t M{\"u}nchen,\\
                       Physik Dept., D-85748 Garching, Germany \\
\vspace{2mm}
        $^b$ Max Planck Institut f{\"u}r Physik (Werner-Heisenberg-Institut),\\
        F{\"o}hringer Ring 6, D-80805 M{\"u}nchen,  Germany }
\vspace{1.5cm}

{\large\bf Abstract}
\end{center}
We study the renormalization of the fermion mixing matrix in the Standard
Model and derive  the constraints that must be satisfied 
to respect  gauge invariance to all orders. 
We demonstrate that 
the prescription based on the {\it on-shell} 
renormalization conditions is not consistent 
with the Ward-Takahashi Identities 
and leads to gauge dependent physical
amplitudes. A simple   scheme is  proposed that
satisfies all theoretical requirements and is very convenient for practical 
calculations.

\noindent


\end{titlepage}
\newpage
Despite  a few  interesting papers have been devoted 
to the fermion mixing renormalization \cite{ms,dono,denner,kniehl}, 
the subject has so far escaped much attention.
This is mostly due to the fact that, as a result of GIM cancelations,
the radiative corrections related to the 
renormalization of the CKM matrix can be made very  small, $O(G_F \,m_q^2)$,
where $m_q$ is the mass of a light quark. They 
 are therefore of little practical
importance in the context of the Standard Model (SM).
Still, the subject has  some conceptual interest in its own, not to mention
the relevance of mixing in many extension of the SM.
In this letter, we  reconsider it from a different  
point of view, which allows us to point out some inconsistency in previous
analyses and to propose an alternative solution.

For definiteness, we concentrate on the case of the CKM matrix renormalization
in the SM.  
A convenient framework to study this issue 
  is  provided by the Ward Takahashi Identities (WTI)
of the theory with  background fields \cite{zuber,abbott,antonio}. 
Indeed, as the diagonalization of the fermion mass matrix is achieved by 
field redefinitions that do not commute with the gauge transformations,
the  CKM elements appear explicitly in the WTI, 
unlike masses and gauge couplings.
This will give us  a strong constraint. 
At the functional level, the WTI which represent 
the $\sigma^+$ generator of $SU(2)_L$ 
are implemented  by the local functional operator ${\cal W}_+$ 
acting on the effective action $\Gamma$
(see \cite{antonio}). For our purposes,
the relevant part of ${\cal W}_+$ is the one which contains the quark
fields:
\be
{\cal W}_+^{quark} = \sum_{u,d}\left[
\bar{\psi}_{u}^L\, V_{ud}^{0} \,\frac{\stackrel{\rightarrow}{
\delta}}{\delta \bar{\psi}^L_d}-
\frac{\stackrel{\leftarrow}{\delta}}{\delta \psi^L_u} \,V_{ud}^{0} 
\,\psi^L_{d} \right].
\ee
Here $V^0$ is the tree level CKM matrix. Upon renormalization, 
the fermionic fields are rescaled by non-diagonal complex 
wave function renormalization (WFR) 
matrices $Z_u$ and $Z_d$;  consequently,  $V^0$ is replaced by
 a renormalized CKM matrix
$V= (Z_u^L)^{-\frac12}\, V^0 \, (Z_d^{L})^{\frac12}$, where 
$Z^L$ is the left-handed
component of the WFR. Expanding $V$ at first order,
 we obtain for the CKM counterterm
\be
\delta V_{ud}= \frac12 \sum_{u'} \delta Z_{u u'}^L V_{u' d} -
 \frac12 \sum_{d'} V_{u d'} \delta Z_{d' d}^L 
\label{wti1}.
\ee
An additional constraint on the WFR
comes from the requirement of unitarity for $V$,
\be
\sum_{d'} V_{u d'} \left(\delta Z_{d' d}^L + \delta Z_{d d'}^{L*}\right)
=\sum_{u'} \left(\delta Z_{u u'}^L + \delta Z_{u' u}^{L*}\right)V_{u'd}
\label{wti2}.
\ee
Notice that this constraint follows also from the request that 
the commutation relations among the operators ${\cal W}_+$, ${\cal W}_-$, and 
${\cal W}_3$ be preserved.
Combining \eqs{wti1} and (\ref{wti2}) we find
\be
\delta V_{ud}=
-\frac14 \left[ \sum_{u'} \left(\delta Z^{L*}_{u'u} - \delta Z^L_{u u'} 
\right) V_{u' d} +
\sum_{d'} V_{u d'} \left(\delta Z^{L}_{d' d} - \delta Z^{L*}_{d d'} 
\right) \right],
\label{deltav}
\ee
where, 
as expected for a unitary matrix, the renormalization of $V$ is expressed in
terms of anti-hermitian matrices.

Any renormalization prescription that preserves the above WTI leads to a
gauge-independent definition of the CKM matrix. To prove this theorem 
we start noticing that, as a consequence of ${\cal W}_+ (\Gamma)=0$,
one has
\be
\partial_{\xi} {\cal W}_+ (\Gamma)=  {\cal W}_+(\partial_{\xi} \Gamma)
+ \sum_{u,d}\left[ 
\bar{\psi}_{u}^L\, \partial_{\xi}V_{ud} \,\frac{\stackrel{\rightarrow}{
\delta}}{\delta \bar{\psi}^L_d}-
\frac{\stackrel{\leftarrow}{\delta}}{\delta \psi^L_u}\partial_{\xi}
\, V_{ud} \,\psi^L_{d} +\dots \right] \Gamma=0
\label{theor}
\ee
where $\xi$ is a gauge parameter
and the ellipses stand for additional  contributions not relevant for us.
On the other hand, the gauge variation of $\Gamma$ is controlled
by a Slavnov-Taylor Identity (STI), $ \partial_{\xi} 
\Gamma= {\cal S} (\partial_\chi \Gamma)$ \cite{zuber,STIgauge},
 where $\chi$ is the 
anti-commuting source of the composite operator generated by the variation 
of $\xi$, and ${\cal S}$ is the Slavnov-Taylor operator \cite{brs}. As
${\cal W}_+$ does not depend on $\chi$ and commutes with ${\cal S}$,
the first term of \equ{theor} vanishes.
The only possibility compatible with the invariance of the theory
(i.e. with \equ{wti1}) is then that all the parameters in the square brackets
of  \equ{theor} are identically zero. From this $\partial_\xi 
V=0$ follows.

An interesting point about \equ{deltav} is that the counterterms for  physical
parameters\footnote{Strictly speaking, 
the CKM element $V_{ud}$ is {\it not}
a physical quantity; physical observables are identified at tree level by 
the Jarlskog reparametrization 
invariants \cite{jarlskog}, which are constructed from
physical amplitudes. This scheme can in principle generalized at higher
orders.}, the CKM matrix elements, 
are given in terms of conventional objects like the WFR
constants. From a rigorous point of view, 
the distinction between  physical and conventional objects 
can be formulated  in terms of the cohomology classes of 
the Slavnov-Taylor operator of the theory \cite{brs}. 
The physical parameters on which the $S$-matrix depends 
are  the coefficients of the cohomology  classes.
For example, in the absence of mixing, 
one  cohomology class is provided by each Yukawa interaction term
of the Lagrangian, which guarantees  that the masses 
 are physical objects.
In the case of mixing, the Yukawa couplings are complex non-diagonal matrices
and can be diagonalized by a redefinition of the fields, that is a finite WFR.
 In higher orders, however, the redefinition 
of fields originated by the diagonalization of the Yukawa matrix inevitably 
mixes with the one generated by the anomalous rescaling of the kinetic 
terms. 
A formal cohomological analysis
\cite{antonio} shows that indeed the off-diagonal
field redefinition contains some physical
parameters, namely  the CKM matrix elements.
On the other hand,
Eqs.~(\ref{wti1}-\ref{deltav}) allow us to disentangle the physical information
related to the diagonalization of the Yukawa matrix (contained in the 
CKM matrix elements and in their counterterms) from the unphysical 
information carried by the $Z$ factors.
These constraints
rely on the gauge invariance of the theory  and, in a consistent
framework, should be  all satisfied.

As we have seen, the definition of the CKM matrix at higher orders is
conveniently expressed in terms of the anti-hermitian component of the 
WFR. It therefore depends on  the scheme chosen for the WFR.
It should be clear, on the other hand, that once 
the counterterm $\delta V$ is calculated through \equ{deltav}, it
can be used independently of the choice of the $\delta Z$ factors, 
because physical amplitudes are {\it independent} 
of the scheme adopted for the WFR.
For example, in practical applications at the level of $S$-matrix, 
it is often convenient to avoid  the rescaling of the fields (i.e. the WFR)
altogether
\cite{sirlin80} and introduce only the LSZ factors for the external fields:
if mixing is present, however, one still has to 
renormalize the mixing parameters.

A number of  renormalization prescriptions for the CKM matrix 
are indeed possible;
 a first convenient option is the  $\ms$ subtraction: 
by definition, assuming gauge invariant mass renormalization
and after adjusting for the possible breaking of chiral invariance, 
 it satisfies the WTI and the STI.
Hence, as a consequence of the above theorem,  it can be guaranteed to 
yield a gauge-independent ultraviolet pole 
 $\delta V^{\ms}$ \cite{pole} to all orders (a proof can also be found
following \cite{wilcz}).
On the other hand, it is well-known that the 
 decoupling of heavy particles is not manifest
 in the $\ms$ scheme. This means that if we work in the framework of
an effective Lagrangian where the heavy fields ($W$ boson and top quark)
are integrated out, the dimension three and four operators
 that mix the quarks yield contributions to the amplitude which are 
not suppressed by the high mass scale. 
Moreover, as noted in \cite{bbbar}, these terms are not even defined in the 
limit of vanishing light quark masses. All this makes the $\ms$ definition
unnatural in the context of effective  Lagrangians.
As the  CKM elements are mostly
determined from low-energy hadronic processes, this  is not very convenient.
Physical amplitudes calculated with an $\ms$ counterterm for the CKM
 would depend on the
renormalization scale (see \cite{mannel} for studies of the scale 
 evolution of the 
CKM matrix) and would contain $O(\alpha)$ corrections proportional to
$(m_i^2+m_j^2)/(m_i^2-m_j^2) $, where $m_{i,j}$ are the  poorly known
light quark masses.

A second   possibility consists in fixing four CKM
elements in terms of four physical amplitudes, e.g. of  the four most
precise experimental processes. This procedure bypasses 
the definition of  the WFR, but destroys the symmetry between the quark
generations and is not practical in higher order calculations.

A third option  is provided by the use of the 
{\it on-shell} renormalization conditions of Ref.\cite{aoki}
 to define the WFR constants and, through \equ{deltav}, the CKM
counterterm, as has been suggested by
  Denner and Sack \cite{denner} and generalized to 
extended models in \cite{kniehl}. 
This approach also implies  decoupling  
in the sense explained before, i.e.  dimension three and four
operators are removed. The renormalization
conditions, which are written  in the $u$ sector 
in terms of the fermionic two-point functions 
$\Gamma_{\bar{u} u'}(\not\!p)$,  
\bea
\bar{u}(m_u) \, \Gamma_{\bar{u} u'}(\not\! p)=0;\ &&
\ \Gamma_{\bar{u} u'}(\not\! p) \,u(m_{u'})=0;\non\\
\bar{u}(m_u)\, \Gamma_{\bar{u} u}(\not\! p)\frac{1}{\not\! p - m_u} =1;
\ &&\ 
\frac{1}{\not\! p - m_u} \Gamma_{\bar{u} u}(\not\! p)\, u(m_{u})=1,
\label{aoki}
\eea
fix the masses of the $u$ quarks and all the $\delta Z_{u u'}$  
by setting them equal to the LSZ factors. Their use in \equ{deltav}
defines a counterterm $\delta V_{ud}$ which makes physical amplitudes finite 
\cite{denner}.

We now  show explicitly
that the latter procedure leads to gauge dependent amplitudes in  one-loop
calculations. To this end, we  consider the decay of a $W$ boson 
into two arbitrary quarks $u$ and $d$ that was studied in  Ref.~\cite{denner}.
We conform to the notation of that paper  and write the one-loop
renormalized amplitude  as
\bea
{\cal{M}}_{ud}&=&V_{ud} \,
{\cal{M}}_0 \left( 1+ \delta_{vert} +\frac{\delta e}{e}
- \frac{\delta s_\smallw}{s_\smallw} + \frac12 \delta Z_\smallw\right) \non\\
&& +   {\cal{M}}_0 \left( \frac12\sum_{u'} \delta Z^{L*}_{u' u} \,V_{u' d} +
\frac12\sum_{d'} V_{u d'} \,\delta Z^L_{d' d} + \delta V_{ud}\right), 
\label{amplit}
\eea
where ${\cal{M}}_0= -g/\sqrt{2}  \ \bar{u}(m_u) \!\not\!\!\epsilon (\mw)
\,a_- \,v(m_d)$, $\bar{u} $ and $v$ are the spinors of
the final-state quarks, $a_\pm=(1\pm\gamma_5)/2$ are the right and left-handed
projectors, and $\epsilon^\mu$ is the polarization vector of the 
$W$ boson.

Let us  now consider the gauge dependence of the individual contributions to 
${\cal{M}}_{ud}$. For our purposes, it is sufficient to consider only 
the $\xi_\smallw$ gauge parameter dependence.
As the total amplitude in \equ{amplit} must be  gauge independent,
 $\partial_{\xi_\smallw}  {\cal{M}}_{ud}=0$. Also the counterterms 
$\delta e $ and $\delta s_\smallw$, as defined in \cite{sirlin80}, do not 
depend on $\xi_\smallw$ at the one-loop level.
On the other hand, 
the gauge variation of the proper vertex $\delta_{vert}$ is not
trivial; it can be studied using the 
STI that governs the gauge dependence of
the Green functions\footnote{A detailed illustration  of 
this  kind of STI \cite{zuber,STIgauge} will be given  in \cite{us}.}.
On the mass shell,
and after contracting with $\epsilon^\mu$ and with the external quark
spinors, one finds the following 
non-linear identity written in terms of 1PI Green functions
\bea
\partial_{\xi_\smallw } \Gamma_{W^+ \bar{u} d} = \sum_{i=1,2} \left[\Gamma^T_{\chi_i\gamma^-
  W^+ } \Gamma_{W^+ \bar{u} d} + \Gamma_{\chi_i\bar{u} \eta_{u'} }
\Gamma_{W^+ \bar{u}' d}   + \Gamma_{W^+ \bar{u} d'}
\Gamma_{\chi_i\bar{\eta}_{d'} d }\right].
\label{sti1}
\eea 
Here $\gamma^\pm_\mu$ and $\eta_{u,d}$ are  the sources
 associated to the BRST variation  of 
$W^\pm_\mu$ and $u,d$ fields and $\chi_{1,2}$ are the sources of the two
independent composite operators generated by the variation of  the two
gauge fixing parameters $\xi_{\smallw,1}$ and $\xi_{\smallw,2}$.  
Following  the common practice, we have 
set $\xi_\smallw=\xi_{\smallw,1}=\xi_{\smallw,2}$.  

At the  one-loop level, \equ{sti1} reduces to
\bea
\partial_{\xi_\smallw } \Gamma^{(1)}_{W^+ \bar{u} d}
= \frac{g}{\sqrt{2}} \sum_{i=1,2} \left[ \Gamma^{T(1)}_{\chi_i\gamma^-
  W^+ } \, V_{ud} +
\sum_{u'}\Gamma_{\chi_i \bar{u} \eta_{u'} }^{(1)}\, V_{u'd}
+\sum_{d'}  V_{ud'} \,\Gamma_{\chi_i\bar{\eta}_{d'} d}^{(1)}
\right] \not\!\epsilon \, a_-\ , 
\label{sti2}
\eea
where the superscript $(1)$ indicates that the proper functions are evaluated
in the one-loop approximation and $T$ that only the transverse component is
considered. All terms are evaluated on the mass-shell of the physical fields.
In a similar way, one may find STI for the two-point functions of the
$W$ boson and of the quarks. 
At one-loop level and adopting the 
standard tadpole renormalization,
 which consists in removing the whole tadpole
amplitude, one finds for the $W$ boson WFR  factor
\be
\left. \partial_{\xi_\smallw} 
\delta Z_\smallw=\partial_{\xi_\smallw } \frac{\partial}{\partial p^2}
\Gamma^{T(1)}_{W^+W^-}(p^2)\right|_{p^2=\mw^2}= 
2 \sum_{i=1,2} \Gamma^{T(1)}_{\chi_i \gamma^-  W^+}(\mw^2). 
\label{sti3}
\ee
The treatment of the quark two-point functions is slightly more involved:
in the case of the $u$ quarks, for instance, we decompose the unrenormalized
self-energy  according to
\be
\Sigma_{u u'}(p)= \Sigma^L_{u u'} \not\! p\,a_- + 
 \Sigma^R_{u u'} \not\! p\,a_+ + 
 \Sigma^S_{u u'} \left( m_u a_- + m_{u'} a_+\right).
\ee
The one-loop STI for $\Sigma_{u u'}$ then reads 
\be
\partial_{\xi_\smallw } \Sigma_{u u'}^{(1)}(p)=
-\sum_{i=1,2} \left[ \Gamma_{\chi_i\bar{u}
  \eta_{u'}}^{(1)}(\not\! p) \left(\not\! p- m_{u'} \right)
+  \left(\not\! p- m_{u} \right)\Gamma_{\chi_i
  \bar{\eta}_{u}u'}^{(1)}(\not\! p)\right].
\label{sti4}
\ee
Splitting
$\Gamma_{\chi_i\bar{u}  \eta_{u'}}^{(1)}(m_u)$  into
its left and  right-handed components, $\Gamma_{\chi_i\bar{u}
\eta_{u'}}^{L,R(1)}$, we find on the mass shell of the $u$ quark  
\bea
m_u \partial_{\xi_\smallw } \left(\Sigma^{L,(1)}_{u u'}+
\Sigma^{S,(1)}_{u u'}\right)&=&
m_{u'} \Gamma_{\chi_i\bar{u}\eta_{u'}}^{L,(1)} -m_u
\Gamma_{\chi_i\bar{u}\eta_{u'}}^{R,(1)}\ ,\non\\
m_u \partial_{\xi_\smallw } \Sigma^{R,(1)}_{u u'} + 
m_{u'}  \partial_{\xi_\smallw } 
\Sigma^{S,(1)}_{u u'}&=& 
m_{u'} \Gamma_{\chi_i\bar{u}\eta_{u'}}^{R,(1)} -m_u
\Gamma_{\chi_i\bar{u}\eta_{u'}}^{L,(1)}\ .
\eea
One can then 
use the above equations 
 in the definition of the {\it on-shell} $\delta Z^L$ 
(see for ex. Eq.~(3.4) of Ref.\cite{denner}) and find 
\be
\partial_{\xi_\smallw } 
\delta Z^{L*}_{u' u}= 2 \sum_{i=1,2} 
 \Gamma_{\chi_i\bar{u}\eta_{u'}}^{R,(1)}(m_u).
\label{sti5}
\ee
An analogous result holds in the $d$ sector.

Inserting \eqs{sti3} and (\ref{sti5}) into \equ{sti2}, we can write the gauge
dependence of the vertex as
\be
V_{ud} \,\partial_{\xi_\smallw}   \,\delta_{vert}=
-\frac12 \,\partial_{\xi_\smallw} 
\left(V_{ud} \,\delta Z_\smallw + \sum_{u'} \delta Z^{L*}_{u' u} \,V_{u' d} +
\sum_{d'} V_{u d'} \,\delta Z^L_{d' d} \right).
\label{sti}\ee
We observe that, as a consequence of the unitarity of $V$, 
in the last two terms of \equ{sti} the dependence on the CKM matrix factorizes;
this is consistent with the fact that the STI  in \equ{sti1}
would be exactly the same,  were  the CKM matrix  diagonal, and (from a
diagrammatical point of view) that 
the one-loop vertices involve one and only one charged quark current. 
 The limit for massless fermions of \equ{sti} agrees with the
expressions reported in \cite{ds92}. Note also that 
 all the gauge dependence of the vertex
is contained in two-point function contributions. 
Using \eqs{amplit} and (\ref{sti}), 
$\partial_{\xi_\smallw}\,{\cal M}_{ud}=0$
 is reduced  to $\partial_{\xi_\smallw}\,\delta V_{ud}=0$.
In other words, we have shown that the gauge-parameter 
dependence introduced by the 
LSZ external field factors (that coincide  with the {\it on-shell} WFR
 constants) is completely absorbed by the proper vertex.
On the other hand, $V_{ud}$ is a parameter of the bare Lagrangian
and  its renormalization condition, in the present framework,
 should preserve the gauge independence.

It may  therefore seem surprising that 
 the counterterm defined on the basis of the  
{\it on-shell} conditions is gauge-dependent.
Indeed, an explicit one-loop calculation yields
for the $\xi_\smallw$ 
dependent part of the WFR in $n$ dimensions
\be
\delta Z_{u' u}^{L,\xi_\smallw  *}=  \frac{ig^2 \mu^{4-n}}{2\mw^2}\sum_{d'} 
V_{u d'} V_{u' d'}^* \int \frac{d^n k}{(2\pi)^n}
\frac{\xi_\smallw \mw^2 -m_u^2+m_{d'}^2 }
{(k^2-\xi_\smallw \mw^2) [(k+p)^2 - m_{d'}^2]}\ ,
\label{deltaZ}
\ee
with $p^2=m_u^2$ and  $u\neq u'$; 
again an analogous expression holds for $\delta Z^L_{d d'}$. 
We see from \equ{deltaZ}
that the $1/(n-4)$ pole  of the anti-hermitian part of 
$\delta Z^{L,\xi_\smallw}$ 
is removed by  GIM cancelations, ensuring that 
the divergence of $\delta V_{ud}$ in \equ{deltav} is gauge independent.
On the contrary, the momentum dependence of the integrand in \equ{deltaZ}
spoils the  GIM cancelations for the finite part of $\delta V_{ud}$
in \equ{deltav},
 which turns then out to be gauge dependent.
We conclude that the $W$-decay amplitude calculated in the {\it on-shell } 
framework of Ref.~\cite{denner} is gauge dependent.
This  can be understood by noting that the finite part of 
the  WFR factors defined {\it on-shell} violates  the WTI, in particular 
it does not satisfy \equ{wti2}.

A convenient and natural  alternative to the prescription of \cite{denner}, 
which maintains decoupling and enhances
the  symmetry among the quark generations, 
can be obtained by imposing the following conditions on the off-diagonal 
two-point functions:
\bea
\left.\Gamma_{\bar{u} u'}(0)=0 \ \ ; \ \ \ \ \
\Gamma_{\bar{u} u'}^L(0)\equiv\frac{\partial}{\partial \! \not \! p}
\,\Gamma_{\bar{u} u'}(\not\! p)\, a_-\,\right|_{\not p=0}=0\, ,
\label{new}
\eea
for $u\neq u'$, and analogously for the $d$ sector.  
These conditions do not fix the diagonal two-point functions.
They also do not fix the off-diagonal hermitian part of the 
right-handed  WFR, which however is finite and suppressed by light quark
 masses.
We choose   to set $\delta{\cal  Z}^{R,H}_{u
   u'}=0$ for $u\neq u'$.
The normalization point, $\not\!p=0$, is the same for all flavors and all
divergences related to the mixing
are correctly subtracted, as they are logarithmic and do not
depend on masses and momenta. In addition,
\eqs{new}  avoid problems
in the treatment of the absorptive parts of the two-point functions
whenever any of the quarks is heavy (typically, in the SM, 
this is the case with the top quark).

At the one-loop level, the antihermitian and hermitian WFR factors
$\delta{\cal  Z}^{L,A}_{u u'}$ and 
$\delta{\cal  Z}^{L,H}_{u u'}$ obtained from \equ{new}
can be expressed in terms of self-energies 
evaluated at zero momentum transfer:
\be
\delta {\cal{Z}}^{L,A}_{u u'}=
\frac{m_u^2+m_{u'}^2}{m_u^2-m_{u'}^2}\left[\Sigma^L_{u u'}(0) +
2\,\Sigma^S_{u u'}(0) \right]; \ \ \ \ \ \ \delta {\cal{Z}}^{L,H}_{u u'}=
-\Sigma^L_{u u'}(0)\,.
\label{newZ}
\ee
It is straightforward to verify that  $\delta{\cal  Z}^{L,H}_{u u'}$
satisfies the condition of \equ{wti2} because it
reduces it to the case of no mixing.
The CKM counterterm obtained by the use
of $\delta{\cal  Z}^{L,A}$ in \equ{deltav} is,   in units of $
g^2/(64 \mw^2\pi^2)$,
\bea
\delta {\cal V}_{ud}&=&  \sum_{u'\neq u}
\sum_{d'} V_{u' d} V_{u d'} V_{u' d'}^* \left\{
 \frac{(m_u^2 +m_{u'}^2)\,m_{d'}^2}{2\,(m_u^2-m_{u'}^2)}
 \left[ \frac3{\epsilon} - 3 \ln \frac{\mw^2}{\mu^2}\right.\right.\ \ \ \non\\
&& \ \ \ \ \left.\left.
-\frac{11-5y}{2(y-1)} - \frac{3(y-2) y}{(y-1)^2} \ln y\right]\right\}
 + \ (u\leftrightarrow d, \ u'\leftrightarrow d')\ ,
\label{newdeltav}
\eea
where $y=m_{d'}^2/\mw^2$ and
$\epsilon=(4-n)/2$. $\delta {\cal{Z}}^{L,A}_{u u'}$ and 
$\delta {\cal V}_{ud}$ are 
 gauge independent, as can be directly seen from \equ{sti4}. 
We stress once more that the use of this counterterm,
based on \equ{new} and consequently on the ${\cal Z}^{L,A}$ factors, 
is {\it independent}  of the choice of WRF in the rest of a calculation, and
corresponds to
just one of the many gauge-invariant definition of the one-loop CKM elements. 
Needless to say, it is always the LSZ procedure (see \eqs{aoki}) to dictate 
the treatment of the external lines. 
 $\delta {\cal V}_{ud}$ 
can therefore be used without modifications in the calculation of
\cite{denner}: in that case, the results for the $W$-decay
 amplitude are gauge independent and differ from \cite{denner}
by gauge dependent $O(G_\mu m_{light}^2)$ terms.

We now consider the consistency of the 
  renormalization conditions of \eqs{new} beyond one-loop.
First, we can show that the \eqs{new}  respect the WTI for the
$W\bar{u} d$-vertex at all orders; this WTI reads 
\bea
(p_u+p_d)_\mu \,\Gamma_{\h{W}^+_\mu \bar{u} d} (p_u,p_d) -\mw \Gamma_{\h{G}^+
\bar{u} d} (p_u,p_d)\ \ \ \ \ \ \ \ && \non\\
-  \frac{g}{\sqrt{2}} \left(\sum_{d'} V_{u d'} 
a_+ \,\Gamma_{\bar{d'} d} (p_d) - \sum_{u'} V_{u' d} 
 \Gamma_{\bar{u} u'} (-p_u) a_-\right)&=&0.
\label{wtivert}
\eea
At the one-loop level and for on-shell amplitudes, 
\equ{wtivert}  holds  even in the case the external Goldstone
boson $G^+$ and the $W^+$ are quantized. 
Differentiating with respect to $p_u$ and   $p_d$
and setting all momenta to zero, one finds that
\be
\sum_{d'} V_{ud'} \,\Gamma^L_{\bar{d}'
 d}(0)- \sum_{u'}  \Gamma^L_{\bar{u} u'}(0)\, V_{u' d}= V_{ud} \, 
G_{\bar{u} d}(0),
\ee
where $G_{\bar{u} d}(0)$ is a convergent term induced by $\Gamma_{\h{G}^+
\bar{u} d}$.
The second condition of \eqs{new} reduces this constraint to the case
of no mixing at all orders. This is the crucial requirement.
Similarly, 
the first condition of \eqs{new}, used in \equ{wtivert} at zero momenta, 
reduces it to the well-known constraint on the renormalization of the Yukawa
coupling in the absence of mixing.

We also investigate the effect of \eqs{new} on the STI:
at the one-loop level, they induce several 
constraints on the renormalization of
$\chi$-dependent Green  functions which appear in
\equ{sti4}, e.g. $\Gamma^{(1)}_{\chi_i \bar{u} \eta_{u'}}$; using them together
with \eqs{new} in the two-loop STI,  the latter can be linearized and  
reduces to its one-loop form at $p=0$. This is a  non-trivial result, as 
the gauge dependence of the two-loop off-diagonal $\Sigma_{u u'}^{(2)}(p)$
can now be written only in terms of two-loop Green functions.
At the one-loop level several simplifications occur: for example
one has $\Gamma^{(1)}_{\chi_i \bar{u} \eta_{u'}}(0) \, a_-=0$, because only the
left-handed source $\eta_{u'}^L$ is involved. Beyond one-loop, 
 $\eta_{u'}^R$ may also contribute, to the effect that  this and analogous  
 simplifications are not granted any longer. 
Consequently, the combination $\Sigma^{L}_{u u'}(0)
+2\,\Sigma^{S}_{u u'}(0)$ is not guaranteed to be gauge invariant at
two or more loops, although
analogous but more complicated gauge-invariant  combinations do exist.
As already mentioned, what is certainly gauge invariant at all orders is  the 
$\ms$ pole of $\delta V$;  we have explicitly verified this property
using the STI.

A related problem concerns  $\Sigma^R$, which  is finite at one-loop 
just because of the GIM mechanism, but it may require a subtraction 
beyond one-loop;  this  would also modify the conditions of \eqs{new}.
Moreover, we   note that
a rigorous  analysis of all the WTI (not only of \equ{wtivert})
at higher orders cannot be performed without
specifying the whole set of normalization conditions.
The previous points show that a complete discussion of a non-$\ms$
renormalization of the mixing matrix 
beyond one-loop becomes extremely complex \cite{us},
and is only partially simplified by  \eqs{new}.

The renormalization conditions of \eqs{new} can be used in  any model
containing Dirac fermion mixing. For example, all has been said applies
directly to the case of lepton mixing, which
is suggested by recent experiments,
if the neutrinos have Dirac masses.
They can also be generalized  to extended models, along
the lines of Refs.~\cite{kniehl}.

In summary, we have reanalyzed the renormalization of the fermion mixing
parameters in the Standard Model. We have reviewed several possibilities 
for the definition of the CKM matrix at higher orders, showing the constraints
they have to satisfy in order to respect gauge invariance. 
In particular, we have demonstrated that 
the prescription based on the {\it on-shell} wave function renormalization 
constants is not consistent with the Ward-Takahashi  Identities 
and leads to gauge dependent physical
amplitudes. We have therefore                 
proposed a simple  scheme that naturally
satisfies all theoretical requirements and is very convenient for practical 
calculations.

\vspace{1.cm}
We thank Bernd Kniehl for suggesting the topic, for a careful
reading of the manuscript and for many useful suggestions and remarks.
We are also grateful to D. Maison for interesting discussions and for reading the
manuscript,  to A. Denner and T. Mannel for communications regarding
Ref.\cite{denner}, to M. Steinhauser for technical help, and to C. Becchi,
G. Degrassi,  and A. Sirlin for stimulating conversations. 

\eject
\end{document}